\documentclass[aps,prl,twocolumn,floatfix,superscriptaddress]{revtex4-2}

\usepackage[utf8]{inputenc}
\usepackage[T1]{fontenc}
\usepackage{amsmath, amssymb, amsxtra}
\usepackage{nccmath, icomma, xspace, bbold}
\usepackage{siunitx}
\DeclareSIUnit\dbkm{dB/km}
\DeclareSIUnit\krec{\ensuremath{\textit{k}_\text{rec}}}
\DeclareSIUnit\Erec{\ensuremath{\textit{E}_\text{rec}}}
\usepackage[short]{datetime}
\usepackage[normalem]{ulem}
\usepackage{color}

\usepackage{nccparskip}
\SetParskip{4pt}
\usepackage{setspace}
\usepackage{enumitem}

\usepackage[british]{babel}
\usepackage{csquotes}

\usepackage{graphicx}
\usepackage{subcaption}
\usepackage{tikz}

\usepackage{xifthen}
\newcommand{\ket}[2][]{ %
	\ifthenelse{\isempty{#1}}%
	{\ensuremath{\xspace\left\vert #2 \right\rangle\xspace}}%
	{\ensuremath{\xspace\left\vert #2 \right\rangle_{\! #1}\xspace}}
	}
\newcommand{\bra}[2][]{ %
	\ifthenelse{\isempty{#1}}%
	{\ensuremath{\xspace\left\langle #2 \right\vert\xspace}}
	{\ensuremath{\xspace\prescript{}{#1}{\!\left\langle #2 \right\vert\xspace}}}
	}
\newcommand{\braket}[3][]{ %
	\ifthenelse{\isempty{#1}}%
	{\ensuremath{\xspace\left\langle #2 \left\vert\right. #3 \right\rangle\xspace}}
	{\ensuremath{\xspace\left\langle #2 \left\vert\right. #3 \right\rangle_{\! #1}\xspace}}
	}
\newcommand{\avg}[2][]{ %
	\ifthenelse{\isempty{#1}}%
	{\ensuremath{\xspace\langle #2 \rangle\xspace}}%
	{\ensuremath{\xspace\langle #2 \rangle_{\! #1}\xspace}}
	}

\newcommand{\krec}{\ensuremath{k_\text{rec}}\xspace}
\newcommand{\Erec}{\ensuremath{E_\text{rec}}\xspace}

\makeatletter
\let\cat@comma@active\@empty
\makeatother

\begin{document}
\title{Generation of fully phase controlled two-photon entangled states}

\author{Ian Ford}

\author{Adrien Amour} 

\author{Matthias Keller}
\affiliation{Department of Physics and Astronomy, University of Sussex, Brighton, BN1 9QH, United Kingdom}

\begin{abstract}

Control over the internal states of trapped ions makes them the ideal system to generate single and two-photon states. Coupling a single ion to an optical cavity enables efficient emission of single photons into a single spatial mode and  grants control over their temporal shape, phase and frequency. 
Using the long coherence time of the ion's internal states and employing a scheme to protect the coherence of the ion-cavity interaction, we demonstrate the generation of a two-photon entangled state with full control over the phase. Initially, ion-photon entanglement is generated. A second photon is subsequently generated, mapping the ion's state onto the second photon. By adjusting the drive field the phase of the entangled state can be fully controlled. 
We implement this scheme in the most resource efficient way by utilizing a single $^{40}\text{Ca}^{+}$ ion coupled to an optical cavity and demonstrate the generation of a two-photon entangled stated with full phase control with a fidelity of up to 82\%.
\end{abstract}

\maketitle

Controlling interaction between matter and light is a fundamental building block for many applications such as quantum networking \cite{kimble2008quantum, duan2010quantum, wehner2018quantum, wei2022towards, azuma2023quantum, Beukers2024RemoteEntanglement}, distributed and blind quantum computation \cite{Monroe2014, Brown2016, Main2025Distributed, Jiang2007DistributedQC, Broadbent2009Universal, Drmota2024Verifiable} and networked sensing and timing \cite{Zhang2021Distributed, Gottesman2012LongerBaseline, Komar2014Quantum, Nichol2022QuantumClocks}.
In particular, multi-photon entangled states are a valuable resource for generating multipartite entanglement across a network \cite{Meignant2019GraphStatesNetworks, Vivoli2019GHZGeneration}, generating photonic graph states  \cite{ Thomas2022GraphStates} or implementing repeaters for long-distance networks \cite{Wallnofer2016TwoDimRepeaters, Krutyanskiy2023TelecomRepeater, Munro2015InsideRepeaters, Azuma2015AllPhotonicRepeaters, Jiang2009QuantumRepeaterEncoding, Simon2007PhotonPairRepeaters, Sangouard2011QuantumRepeaters, Langenfeld2021QuantumRepeaterQKD}.
The consecutive emission of photons with real-time control of the emission process can be employed to implement a heralded quantum memory \cite{Bussieres2013OpticalMemories, Tanji2009HeraldedMemory} and enables feed-forward schemes in linear optics quantum computing \cite{Knill2001LinearOpticsQC, Kok2007LOQCReview, Wootton2012QuantumMemories}.
Due to the long coherence time, and good control over their internal quantum states, ions are prime candidates for implementing and demonstrating schemes for many of these applications. Recently, important aspects of distributed and blind QC have been demonstrated \cite{Drmota2024Verifiable, Main2025Distributed} using spontaneously emitted photons. Employing ions coupled to optical cavities promises enhancement in efficiency and improved control of the photonic emission process \cite{PhysRevLett.124.013602, Walker2020, Keller2022TimeBinPhotons}. 
In recent years, ion-cavity systems have been used to demonstrate several crucial building blocks for quantum networking \cite{stute2012, stute2013, Krutyanskiy2023Entanglement230m}. \\
In this paper we present the first demonstration of two-photon entangled states generated with a single ion coupled to an optical cavity. Using two consecutive cavity-assisted Raman transitions (similar to \cite{Walker2020}), we generate two, polarization-entangled photons. Because the intermediate state is an ion-photon entangled state, we can control the phase of the two-photon state by manipulating the drive laser parameters of the second photon generation stage.
Providing full control over the two photon emission processes and highly indistinguishable photons, this scheme is relevant beyond quantum networking for applications such as quantum memories and feed-forward schemes in linear optics quantum computers. 

\begin{figure}[htbp]

    % First row
    \begin{subfigure}{0.45\textwidth}
        \begin{tikzpicture}
            \node[anchor=south west, inner sep=0] (image) at (0,0) 
            {\includegraphics[width=\textwidth]{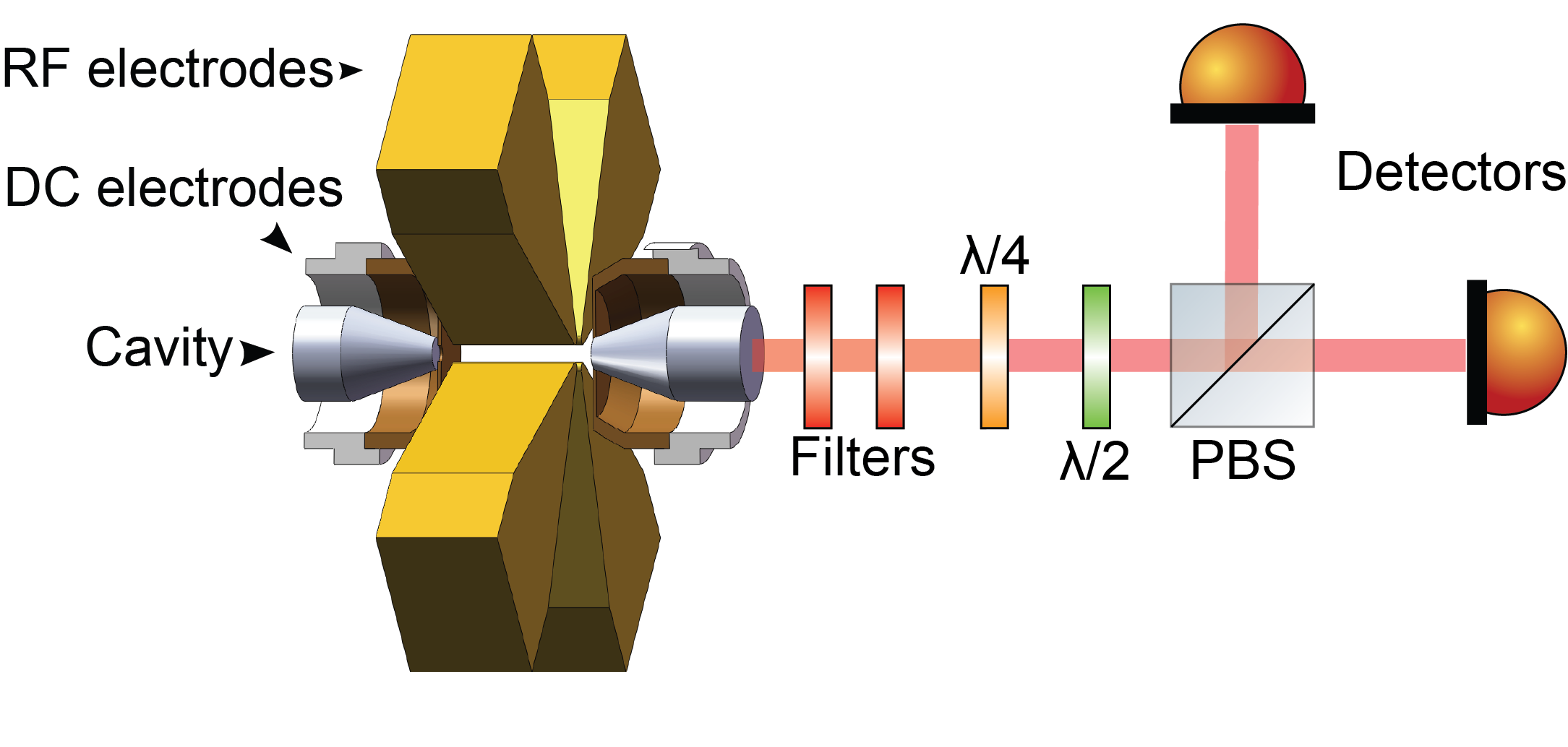}};
            %\node[anchor=north west, font=\small, xshift=4pt, yshift=15pt] at (image.north west){(a)};
        \end{tikzpicture}
    \end{subfigure}
    \caption{Single $^{40}\text{Ca}^{+}$-ions are trapped in a linear Paul trap at the antinode of an optical cavity. The cavity mirrors are placed within the DC end-cap electrodes with an ion-cavity coupling $g_0=2\pi \times0.76$ MHz. The generated photons are sent through two band-pass filters to remove the $894$ nm cavity lock light and subsequently through a pair of waveplates and a PBS whose outputs are each coupled to a single photon detector.}
    \label{fig:Trap Setup}
\end{figure}

\begin{figure*}[htbp]
\vspace*{-2cm}
\begin{center}
 \includegraphics[width=\linewidth]{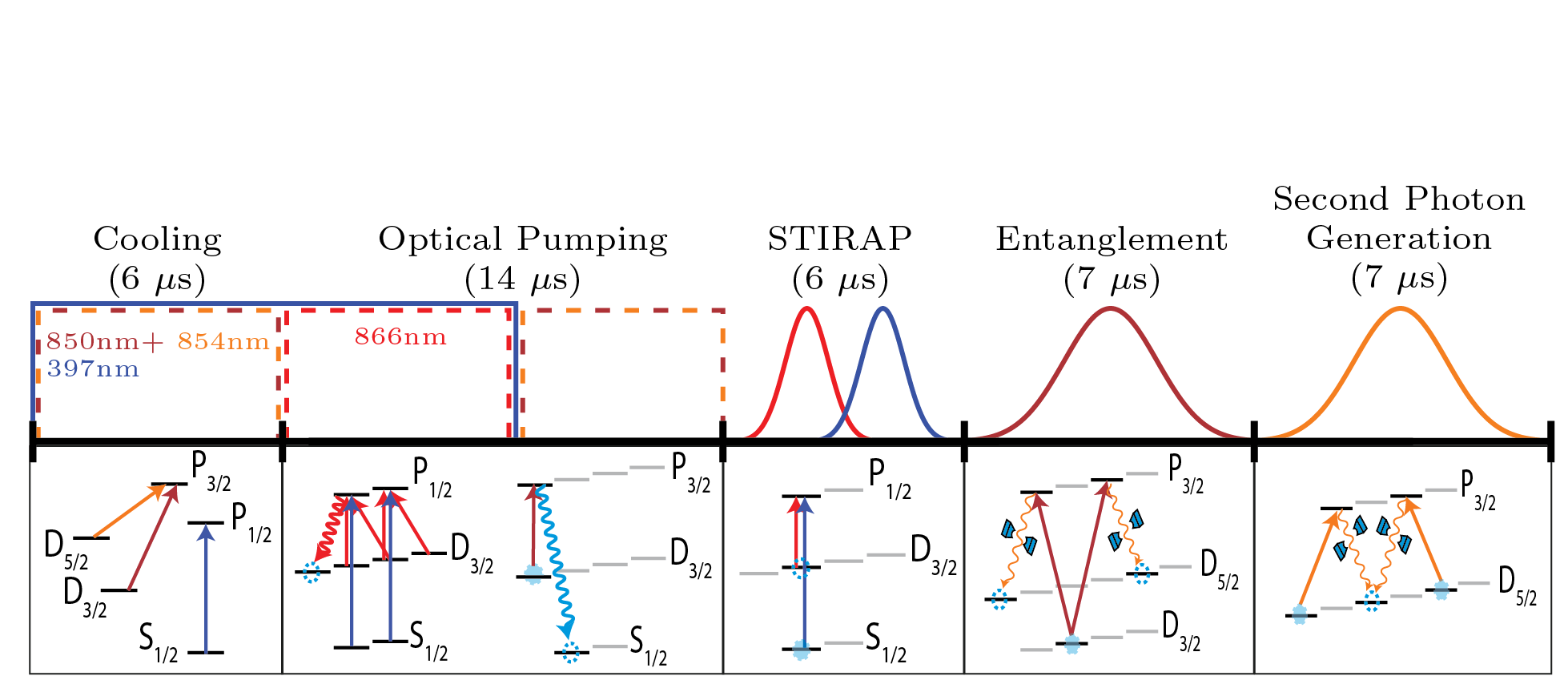}
 \caption{Experimental pulse sequence for creating entangled photon pairs. The ion is Doppler cooled for $6$ $\mathrm{ \mu s}$ before being optically pumped into the S$_{1/2}$,$m_j=-1/2$ state with >95\% efficiency. From this state the ion is transferred to the D$_{3/2}$,$m_j=-1/2$ state via STIRAP. A bi-chromatic 850 nm pulse produces an ion-photon entangled state. A second bi-chromatic drive at 854 nm generates a second photon producing a two-photon entangled state.
   \label{fig:seq}}
   \end{center}
\end{figure*}

\begin{figure}[bp]
    % First row
    \begin{subfigure}{0.45\textwidth}
        \begin{tikzpicture}
            \node[anchor=south west, inner sep=0] (image) at (0,0) 
            {\includegraphics[width=\textwidth]{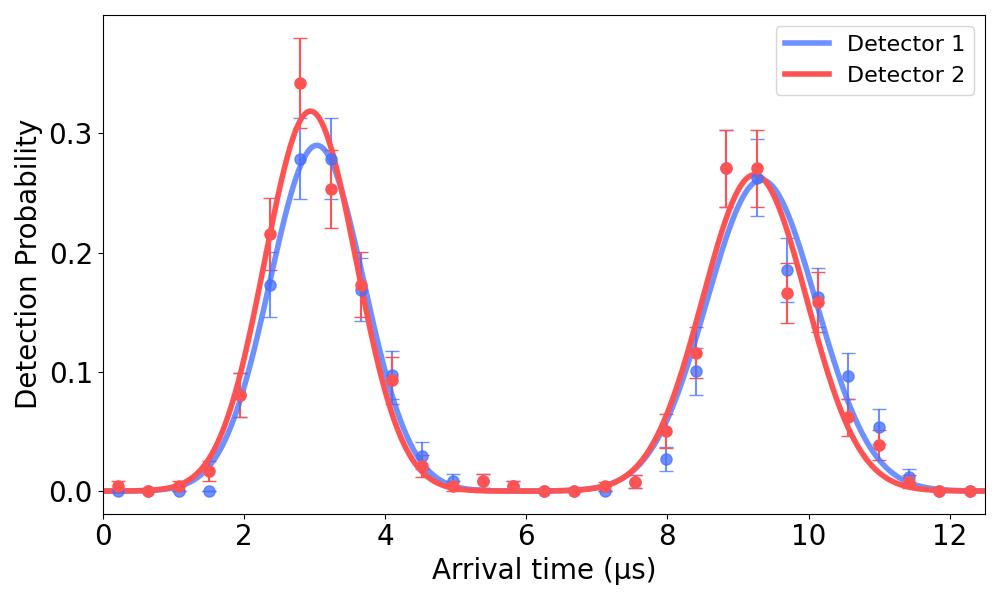}};
            %\node[anchor=north west, font=\small, xshift=4pt, yshift=15pt] at (image.north west){(a)};
        \end{tikzpicture}
    \end{subfigure}
    \caption{Photon detection probability for the two polarization channels for measured entangled pairs. Each photon area is normalized to 1 across both polarizations.}
    \label{fig:Photon Time Trace}
\end{figure}

The ion trap used in this experiment is shown in Figure \ref{fig:Trap Setup}, and is described in more detail in \cite{Walker2018LongDistanceQFC}. A single $^{40}$Ca$^+$ ion is trapped in a linear Paul trap and coupled to an optical cavity, which is aligned collinear to the trap axis. The cavity has a length of $5.75$ mm and is formed by mirrors with a radius of curvature of 25 mm that are recessed within the endcaps of the trap.
Both mirrors are mounted on ring piezo actuators to stabilize the cavity length and to translate the cavity with respect to the ion to maximize the ion-cavity coupling \cite{Multi_Ion}.
The cavity coupling at the antinode is $g_0=2\pi \times 0.76$ MHz and the cavity decay rate is $\kappa = 2\pi \times 0.27 $ MHz. 
The cavity length is stabilized using a laser at 894 nm  referenced to the cesium atomic D$_1$ line.
A magnetic field of $8.25$ G along the cavity axis provides the quantization axis as well as a large Zeeman splitting between levels to allow for addressing individual transitions within the ion. With this magnetic field along the cavity axis, the cavity couples solely to $\sigma^+$ and $\sigma^-$ transitions of the ion. The cavity is tuned to the zero-field $P_{3/2}- \mathrm{D}_{5/2}$ atomic transition at $\sim$854 nm.
 
To generate the two-photon entangled state, we use the following sequence (see Figure \ref{fig:seq}). The ion is initially cooled on the S$_{1/2}-$P$_{1/2}$ transition with lasers at 850 nm and 854 nm used to repump from the metastable $\mathrm{D}_{3/2}$ and $\mathrm{D}_{5/2}$ states. Subsequently, the ion is optically pumped into the $\mathrm{D}_{3/2},\: m_j=-3/2$ state using a $\pi + \sigma^-$ polarized 866 nm laser, followed by a second optical pumping stage using lasers at  850 nm and 854 nm to prepare the ion in the $S_{1/2},\:m_j=-1/2$ state. The final initialization step is a  STIRAP pulse to transfer the population into the $\mathrm{D}_{3/2},\:m_j=-1/2$ level using $\pi$ polarized lasers at 397 nm and 866 nm. Utilizing the metastable state as the initial state for entanglement significantly improves indistinguishability of the generated photons  as shown in \cite{Walker2020}.

\begin{figure*}[htbp]
%    \centering

    % First row
    \begin{subfigure}{0.49\textwidth}
        \begin{tikzpicture}
            \node[anchor=south west, inner sep=0] (image) at (0,0) 
            {\includegraphics[width=\textwidth]{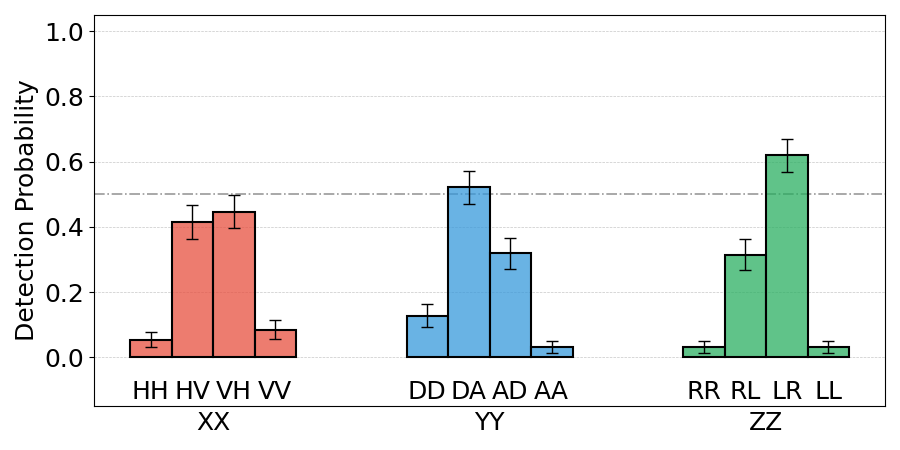}};
            \node[anchor=north west, font=\small, xshift=4pt, yshift=15pt] at (image.north west){(a)};
        \end{tikzpicture}
        \label{fig:fidelity1}
    \end{subfigure}
    \hfill
    \begin{subfigure}{0.49\textwidth}
        \begin{tikzpicture}
            \node[anchor=south west, inner sep=0] (image) at (0,0) {\includegraphics[width=\textwidth]{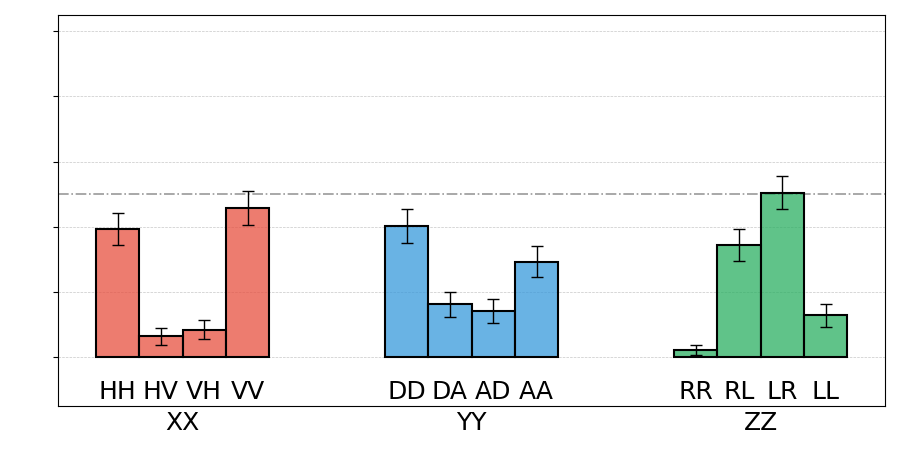}};
            \node[anchor=north west, font=\small, xshift=4pt, yshift=15pt] at (image.north west){(b)};
        \end{tikzpicture}
        \label{fig:fidelity2}
    \end{subfigure}

    % Second row
    %\vskip\baselineskip
    \begin{subfigure}{0.49\textwidth}
        \raisebox{12.5pt}{
        \begin{tikzpicture}
            \node[anchor=south west, inner sep=0] (image) at (0,0)
            {\includegraphics[width=\textwidth]{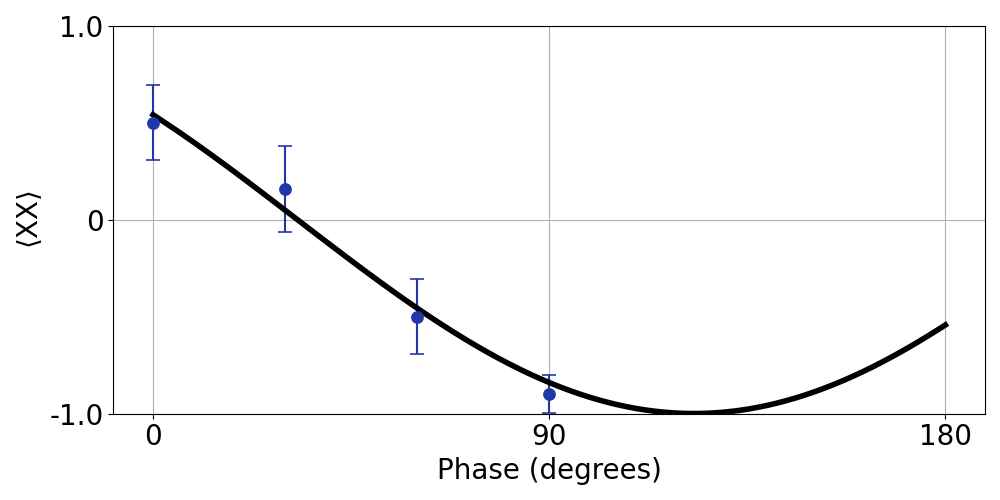}};
            \node[anchor=north west, font=\small, xshift=4pt, yshift=15pt] at (image.north west){(c)};
            \end{tikzpicture}
            }
        \label{fig:phase1}
    \end{subfigure}
    \begin{subfigure}{0.49\textwidth}
        \raisebox{10pt}{
        \begin{tikzpicture}
            \node[anchor=south west, inner sep=0] (image) at (0,0) {\includegraphics[width=\textwidth]{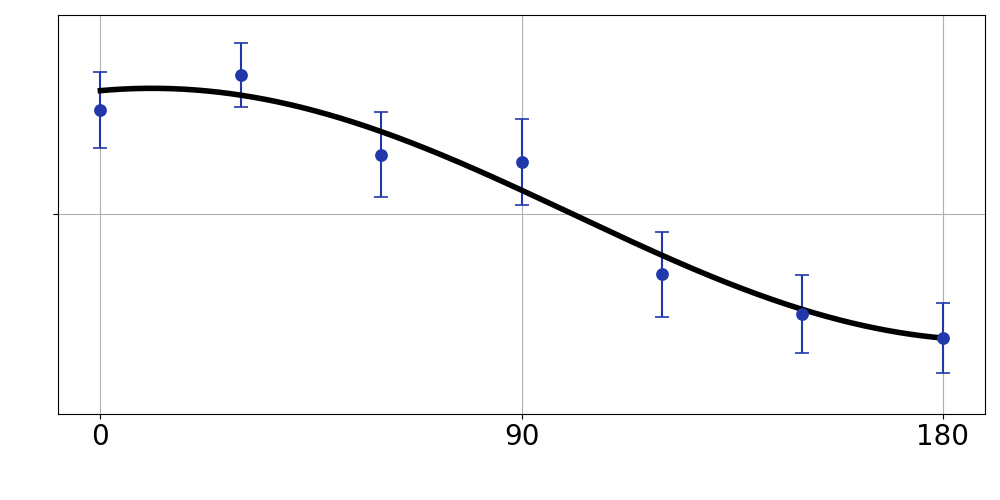}};
            \node[anchor=north west, font=\small, xshift=4pt, yshift=15pt] at (image.north west){(d)};
        \end{tikzpicture}
        }
        \label{fig:phase2}
    \end{subfigure}
    \caption{Detector correlation outcomes for all three measurement bases (H/V, A/D, R/L), used to calculate fidelity of the created (a) $\ket{\Psi^-}$ and (b) $\ket{\Psi^+}$ Bell states, showing successful creation of entangled photon pair with fidelity of 0.82 and 0.70 respectively. The gray dashed lines mark the measurement efficiency for a perfect Bell state. (c) and (d) showing contrast measurement in X-basis as phase is scanned to find the optimum. Solid lines are fitted sin curves of form $A \sin(x+\phi)$.}
    \label{fig:main result}
\end{figure*}
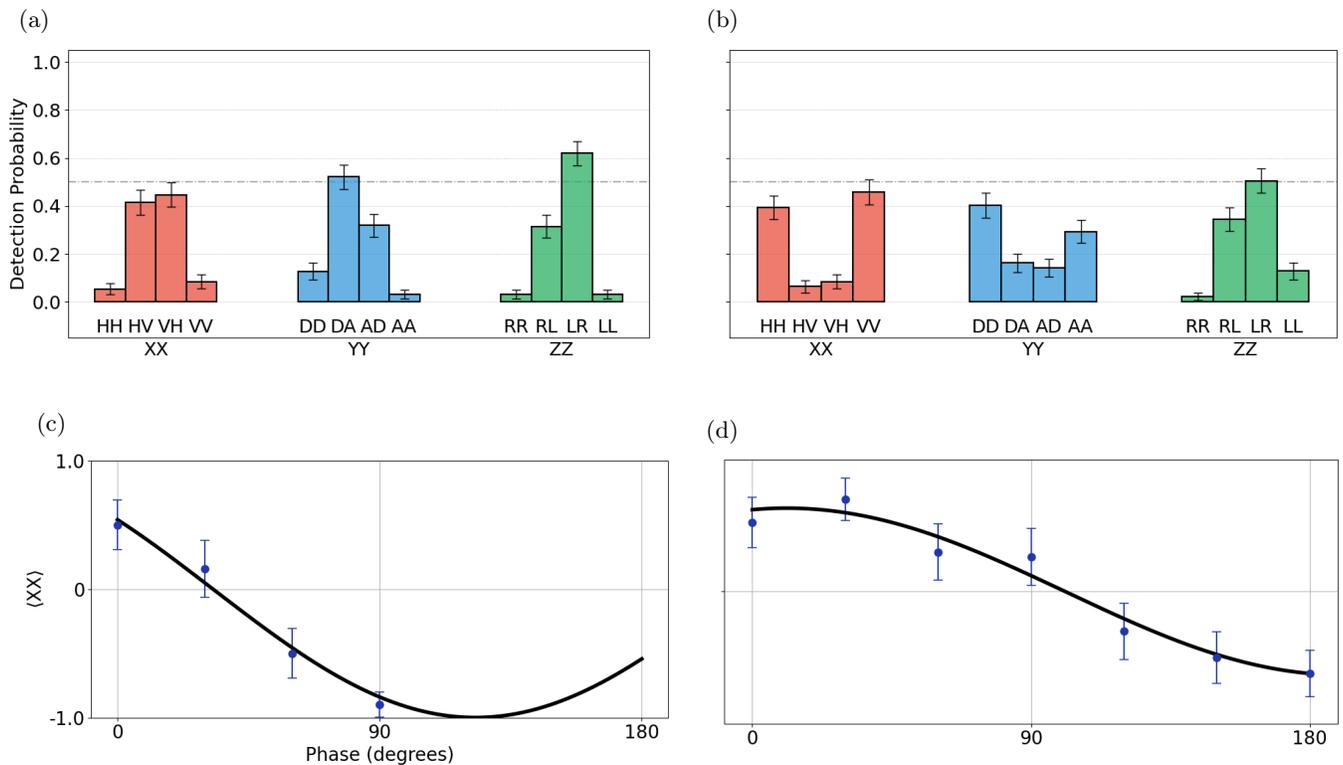

A single 854 nm photon is subsequently created by a bi-chromatically driven cavity-assisted Raman transition with a $\sigma^+ + \sigma^-$ polarized laser at 850 nm. Each tone of the bi-chromatic drive is tuned such that both are on Raman resonance with a transition to a different Zeeman sublevel of the D$_{5/2}$ manifold, each of which produces a different polarization of photon, thereby entangling the ion with the emitted photon. 
This ideally produces the maximally entangled state:
\begin{equation}
    \ket{\Psi}_{\mathrm{ion-photon}} = \frac{1}{\sqrt{2}}(\ket{\mathrm{D}, \sigma^+} - \ket{\mathrm{D}', \sigma^-})
    \label{eq:ion-photon}
\end{equation}
 where $\ket{\mathrm{D}}$ and $\ket{\mathrm{D}'}$ are the atomic states $\mathrm{D}_{5/2},\:m_j=-5/2$ and $\mathrm{D}_{5/2},\:m_j=+3/2$ respectively. 
A second bichromatic cavity assisted Raman transition with 854 nm drive laser maps the state of the ion onto a second photon, hence creating the final entangled photon-photon state:
\begin{equation}
    \ket{\Psi}_{\mathrm{photon-photon}} = \frac{1}{\sqrt{2}}(\ket{\sigma^-, \sigma^+} + e^{i(\phi+\phi_m)} \ket{\sigma^+, \sigma^-})
    \label{eq:photon-photon}
\end{equation}
where all contributions to the superposition phase from atomic structure and deterministic quantum dynamics are included in $\phi$.
The additional phase $\phi_m$ is determined by the relative phases of the fields used in the ion-photon entanglement generation stage and the ion-photon state map stage, as the atomic bases in equation \ref{eq:ion-photon} are determined by the entangling fields.
The two photons have a separation of $6$ $\mu$s with negligible temporal overlap.
The cavity emission is sent through a motorized quarter and subsequent half waveplate to change the polarization basis of the photons. A Wollaston prism (PBS) is employed to measure the polarization, with each output coupled to a superconducting nanowire single photon detector (SNSPD). The PBS has an extinction ratio of $>$15000:1 for both outputs.
In order to measure the photon-photon entanglement fidelity, we produce the $\ket{\Psi^{\pm}}$ Bell states, for which the fidelities are given by: 
\begin{equation}
    \mathcal{F}_{\pm} = \frac{1}{4}(1 \pm \langle XX \rangle \pm \langle YY \rangle - \langle ZZ \rangle),
    \label{eq:Fidelity}
\end{equation}
with $\langle XX \rangle$, $\langle YY \rangle$ and $\langle ZZ \rangle$ being the 
expectation values for the two-photon Pauli operators.
For the pure state shown in equation \ref{eq:photon-photon}, $\langle XX\rangle$ can be expressed as:
\begin{equation}
    \langle XX\rangle=\cos (\phi+\phi_m),
\end{equation}
and hence 
\begin{equation}
   \langle XX\rangle=\begin{cases}
       +1 \:\;\text{for}\:\; \ket{\Psi^{+}}\\
       -1 \:\;\text{for}\:\; \ket{\Psi^{-}}.
   \end{cases}
\end{equation}
Thus, by measuring $\langle XX\rangle$ whilst tuning the relative phase between the two drive beams for the second photon generation step, we can tune the superposition phase to the desired target states. During the tuning process, oscillations in value of $\langle XX\rangle$ were observed, demonstrating control of the phase of the entangled state, with the contrast of the oscillations serving as an indication of the degree of coherence of the entanglement.

Based on this, the experimental sequence proceeded as follows:
first the 850 nm bichromatic drive beam powers and detunings were optimized to generate a photon with polarizations $\sigma^+$ and $\sigma^-$ with equal probability. This was then repeated for the second photon with 854 nm laser. Subsequently the RF phase on one of the tones in the bichromatic drive used for ion-photon state map was tuned whilst measuring $\langle XX\rangle$ to achieve the desired phase, generating $\ket{\Psi^{-}}$ or $\ket{\Psi^{+}}$ states (see Figure \ref{fig:main result}(c-d)). To measure the fidelity, 100 two-photon events were measured per basis. Two-photon events were rejected by post selection based on arrival time to remove any events where a photon was measured outside of its timing window, in order to minimize the effects of detector dark counts. 

The measurement results in each basis for the $\ket{\Psi^{-}}$ and $\ket{\Psi^{+}}$ states are shown in Figure \ref{fig:main result}. For $\ket{\Psi^{-}}$, 94 two-photon events contributed to the measurement of $\langle XX\rangle=-0.72 \pm 0.07$, 94 two-photon events to the measurement of $\langle YY\rangle=-0.68 \pm 0.08$ and 92 two-photon events to the measurement of $\langle ZZ\rangle= -0.87 \pm 0.05$ corresponding to a fidelity of $\mathcal{F}= 0.82 \pm0.03$.
For $\ket{\Psi^{+}}$, 94 two-photon events contributed to the measurement of $\langle XX\rangle=0.70 \pm 0.07$, 92 two-photon events to the measurement of $\langle YY\rangle=0.39 \pm 0.10$ and 93 two-photon to the measurement of $\langle ZZ\rangle= -0.70 \pm 0.07$ corresponding to a fidelity of $\mathcal{F}=0.70 \pm 0.04$. 

The main contribution to the infidelities comes from uncertainty in the Zeeman splitting of the two atomic states.
Any discrepancy between the Zeeman splitting and frequency splitting in the drive fields used for ion-photon entanglement and state mapping leads to an energy difference between the two polarizations of photons.
A consequence of this energy difference is a phase accumulation between the detection of the two photons at rate $\Delta$, where $\Delta$ is the energy difference between the two photons  \cite{doi:10.1126/science.1143835,stute2012}. 
Using our phase control, we can compensate for the average phase accumulation, but due to the finite size of the photon wave-packet there is a random (near Gaussian) distribution in the separation of arrival times and hence a random distribution of superposition phases.
If we let $\tau$ represent the difference between the average precession time and the precession time for a given two photon event, this infidelity can be estimated using the analytic formula,
\begin{equation}
    \epsilon = 1-\frac{1}{2}\int^\infty_{-\infty} P(\tau)\left(1+\cos(\Delta \tau)\right) d\tau
\end{equation}
for probability distribution $P(\tau)$.
Because we are not post-selecting with respect to the detection time of the two photons (in contrast to \cite{doi:10.1126/science.1143835}), we have to integrate over all possible detection delays. In our experiment we measured the Zeeman splitting using the single photon emission spectrum to better than 200 kHz.
We fit $P(\tau)$ using the photon arrival times and found that this effect can contribute an error of up to 25\% to for a worst case measurement of the Zeeman splitting.

An additional infidelity arises during the state-map procedure. Due to different Clebsch-Gordan coefficients of the two cavity-assisted Raman transitions, an asymmetry in Rabi frequency is required for equal transition probability. This leads to differential AC stark shifts, and hence a time-dependent relative phase accumulation during the generation of the second photon. Simulations of the state map procedure of the full 18-level ion-cavity system suggest that this is accounts for a fidelity drop of $\sim$5\% for optimized process efficiency.
By decreasing the Rabi frequency of the drive fields, it is possible to improve the fidelity at the expense of efficiency. 
For example, an ion-photon state map with 2/3 optimal efficiency was found to improve the error to just $1\%$.
In practice, this would require a time intensive experimental calibration.

Other contributions to the infidelity budget are magnetic field drift, laser drift and amplitude noise. However simulations indicated that their contributions to the infidelity budget are small ($\sim$1\%). 

% conclusion
In this publication, we demonstrate the generation of fully phase-controlled two-photon entangled states from an emitter-cavity system for the first time. We have demonstrated the generation of $\ket{\Psi^\pm}$ states with fidelities of above 70\%. The theoretical limit of the fidelity for our experimental parameters is 95\%, limited by the differential AC Stark shift in the photon generation processes due to the asymmetry in our scheme. By decreasing the drive laser intensities fidelities of >99\% are feasible but this results in a 1/3 reduction of the state map efficiency and increased photon pulse length.
The scheme presented here offers the possibility to generate any arbitrary two-photon state by  interleaving the photon generating steps with a coherent rotation between the two atomic states. 
Full control over the two-photon state's phase enables new schemes in linear-optics quantum computing, quantum memories, and quantum networking.

\begin{acknowledgments}
We gratefully acknowledge support from EPSRC through the UK Quantum Technology Hub: QCS - Quantum Computation \& Simulation EP/T001062/1.\\
\end{acknowledgments}

M.K. conceived the scheme. A.A. and I.F implemented and optimized the experimental setup, performed the experiment and analyzed the data. All authors contributed to finalizing the manuscript.

\bibliography{Bibliography}{}

\end{document}